# A Systematic Approach for MRI Brain Tumor Localization, and Segmentation using Deep Learning and Active Contouring

Shanaka Ramesh Gunasekara[1]*, H.N.T.K.Kaldera[2], Maheshi B. Dissanayake[3]*

[123]Dept. Of Electrical and ElectronicEngineering,Faculty of Engineering, University ofPeradeniya, 20400, Sri Lanka.

Correspondence should be addressed to Shanaka R.Gunasekara; shanakag@eng.pdn.ac.lk ,orMaheshi B. Dissanayake; maheshid@eng.pdn.ac.lk

## Abstract

One of the main requirements of tumor extraction is the annotation and segmentation of tumor boundaries correctly. For this purpose, we present a threefold deep learning architecture. First classifiers are implemented with a deep convolutional neural network(CNN) andsecond a region-based convolutional neural network (R-CNN) is performed on the classified images to localize the tumor regions of interest. As the third and final stage, the concentratedtumor boundary is contoured for the segmentation process by using the Chan-Vesesegmentation algorithm. As the typical edge detection algorithms based on gradients of pixel intensity tend to fail in the medical image segmentation process, an active contour algorithm defined with the level set function is proposed. Specifically, Chan-Vese algorithm was applied to detect the tumor boundaries for the segmentation process. To evaluate the performance of the overall system, Dice Score,Rand Index (RI), Variation of Information (VOI), Global Consistency Error (GCE), Boundary Displacement Error (BDE), Mean absolute error (MAE), and Peak Signal to Noise Ratio (PSNR) werecalculated by comparing the segmented boundary area which is the final output of the proposed, against the demarcations of the subject specialists which is the gold standard. Overall performance of the proposed architecture for both glioma and meningioma segmentation is with average dice score of 0.92, (also, with RI of 0.9936, VOI of 0.0301, GCE of 0.004, BDE of 2.099, PSNR of 77.076 and MAE of 52.946), pointing to high reliability of the proposed architecture.

## Introduction

Medical image classification and segmentation is a field, where Deep Learning can make a huge impact with promising results. It facilitates automation of non-invasive imaging based diagnosis. Interestingly, computer aided brain tumor diagnosis has effectively utilized the advances in medical image processing in the past and has opened up many promising research activities in the domain of deep learning, with the expectation of developing entirely computerized automatic accurate diagnostic systems for physicians.

A braintumor is a mass or growth of abnormal cells in the brain which might be cancerous (malignant) or non-cancerous (benign). The early, comprehensive diagnosis and proper treatments are essential for a patient's survival in brain tumor management. During the past decades, more than 120 types of brain tumors were discovered by medical scientists. These brain tumors can be broadly categorized into two main groups, namely; primary brain tumors, which originate in the brain itself and secondary deposits in the brain, where primary tumoriselsewhere in the body[1]. Typically, non-invasive medical imaging techniques such as Computer Tomography (CT) and Magnetic Resonance Imaging (MRI) are favoured as brain tumor identification tools at the initial stages, over incursion invasive procedures like tissue biopsies [2][3]. Authors in [4] found that CT, MRI and Positron Emission Tomography (PET) usage has increased by 7.8% , 10% and 57% respectively during the period of 1996-2010.Furthermore, as ofhealthcare resource statisticsof EU for 2020[5], the EU Member States have shown widespread increase in the availability of medicalimaging technology and equipment for diagnosis in the recent decades. Moreover, according to[6], the overall employment of radiologic and MRI technologists grows faster, than the average for all occupations in USA. All these findings confirm that medical image based diagnosis is favoured in modern healthcare system.

Image classification and segmentation have shown rapid growth during the past two decades with the introduction of machine learning and computer vision techniques. Deep learning has found applications in medical imaging as in, identifying local anatomical characters, detecting organs and body parts, and identifying cells of different shapes and sizes[2]. A review of the related works [4] shows that a considerable portion of the latest research on image analysis uses deep convolutional neural network (DCNN) for both image classification and segmentation[7]. Multimodal Brain Tumor Segmentation (BRATS) Challengeis the main competition on brain tumor classification which is organized by the Perelman School of Medicine at the University of Pennsylvania, Centre for Biomedical Image Computing & Analytics (CBICA) from 2012 onwards. The BRATS challenge focuses on automating the brain tumour detection and the survival rate estimation techniques and algorithms. Each year the dataset is updated and the overall performance of the proposed algorithms have shown a tremendous improvement over time. On the whole, the accuracy of the algorithms proposed using BRATS dataset falls around 90% [8][9][10].Some of these algorithms were developed using classical CNN architecture whereas some are developed using improved CNN algorithms like U- net[11],super pixel-based extremely randomized trees[12].

Another popular and publicly available brain tumor dataset is the Figshare MRI dataset[13][14] which is the dataset employed in this paper. Due to the easy accessibility and the readily availability, Figshare MRI brain tumour dataset also has been used in many Brain tumor classification and segmentation related research [15][16][17][18]. The dataset which was initiate in 2015 and last updated in 2017[13][16], carries an average classificationaccuracy in the range of 90-95%[14][16][19], [20]. Authors in [16] achieved classification average 95% accuracy by using a modified CNN architecture while authors in [15]achieved around 96% accuracy with an automatic content-based image retrieval (CBIR) system. A deep network was enhanced by employing cross channel normalization (CCN) and parametric rectified linear unit (PRELU) in [18] for brain tumor segmentation.

Although convolutional neural network (CNN) has achieved a considerable performance gain in the medical image classification[14], and segmentation tasks, it is associated with a significant increase in the computational cost, especially when high resolution images are analysed. In general, an object detection algorithm draws a bounding box around the object of interest in the image. In a classical automated system, this can be achieved by using a typical convolutional network, followed by a fully connected layer. However, when the number of objects required to be detected is not a fixed number, it is difficult to proceed with the above approach as it requires defining the length of the fully connected layer at the initial design stage.

In the previous work by the authors, a region proposal algorithm is proposed to address the problem of selecting a random number of objects in a single region [21][22].In the proposed method, instead of searching the entire image for number of objects, the algorithm search for objects in several selective areas of the image, while treating each sub-region as an independent sub-image. In [21],a fully autonomous learning algorithm was constructed using Region-based Faster Convolutional Neural Network (Faster R-CNN) to localize the meningioma tumor regions in MRI. Once the tumor is segmented, Prewitt and Sobel edge detection algorithms are applied to the segmentation output, with the expectation of detecting the exact tumor boundary. Both of these techniques compute an approximate tumor boundary using the gradient intensity function of the image [23]. As MRIs on the whole consist of Rician noise, and edges are not defined only by gradient, these algorithms underperform in this task.In practice, the effectiveness of the developed deep learning models to make informed decisions are evaluated through accuracy and system loss. Going beyond simple accuracy, standard mathematical objective parameters such as precision, recall, dice score are utilised to choose the best model for the given problem. Furthermore, graphical representations such as confusion matrix and receiver operation characteristic (ROC) too are utilised to evaluate the performance of the deep learning model. A Confusion matrix is a two dimensional matrix which summarises the performance of the classification algorithm. One dimension of the matrix represents true classes of an object while the other represents the class that classifier predicts[24][25]. The ROC curve is also a two dimensional plot which illustrate how well a classifier system works as a discrimination cut-off value is changed over the range of predictor variable[25].

In the research presented, we outline an automated systematic approach to classify, segment and extract the exact tumour boundary from MRI images. The key contribution of this research can be summarised as follows;

1. We present a simplified CNN architecture based on small number oflayers and faster R-CNN, for the classification of axial MRI into glioma, and meningioma brain tumors, and produce a bounding box of the tumor with 94% of accuracy confidence level [21][22]. One of the key challenges in medical image analysis is the scarcity of labelled data. Hence, in this research we specifically focus on applying R-CNN based tumorlocalisation for ascenario with lower number of annotated data. Specifically, we have used a dataset with less than 500axial MRI images for our research.Furthermore, we kept our network simple, to reduce the total number of trainable parameters.
2. The exaction of the exact tumor boundary was performed by means of unsupervised active contour detection method, developed by means of Chan and Vese[26] algorithm.One of the inherent drawbacks of using active contour algorithms for segmentation is the requirement of initial search area. Without correct initial search area, the algorithm is at the risk of segmenting unwanted regions in the brain MRI, such as orbital cavity and lateral ventricles. In this research, we used tumor bounding box coordinates obtained at faster R-CNN based localisation step as the initial search area for Chan-Vesealgorithm to obtain a refined segmentation of the tumor.This in turn reduces the search area of the Chan-Vese algorithm and helps the algorithm to converge to an accurate boundary in a shorter time frame.The resultant segmented boundary carries a high accuracy according to objective quality evaluation measures, compared to the state of the art.

In this article we were able to present an end to end complete systematic approach for both meningioma and glioma tumor detection and segmentation using MRI. The complete system comprises of 3 main sub systems; brain tumor classification using a simple CNN algorithm, Faster R-CNN based network for tumor localization, and finally Chan-Vese algorithm for exact tumor segmentation.All three algorithms were connected in a cascade manner, with the final deliverable as the exact tumor boundary for segmentation purpose for any given axial brain MRI. The rest of the paper is organized as follows. In the section II we present a summarized over view of the proposed frame work. Section III, briefly outlines the theoretical background of the research. The methodology adopted for the dataset preparation, classification, segmentation and contouring is presented in Section IV. Section V presents the performance analysis, using objective matrices, whereas Section VIdiscusses and compares our proposed architectureagainst the existing works in the literature. Finally Section VII concludes the paper.

## Architecture of the ProposedAlgorithm

In this paper, we propose a threefold completearchitecture to classify and segment brain tumors using T1 weighted MRI sequence. The proposed system architecture consists of three cascaded algorithms; namely in the orderof application, Convolutional neural network for classification, Faster R-CNN network for tumor localization, Chan-Vese algorithm for precisetumor segmentation.The flow diagram of the complete architecture is illustrated in Figure 1.

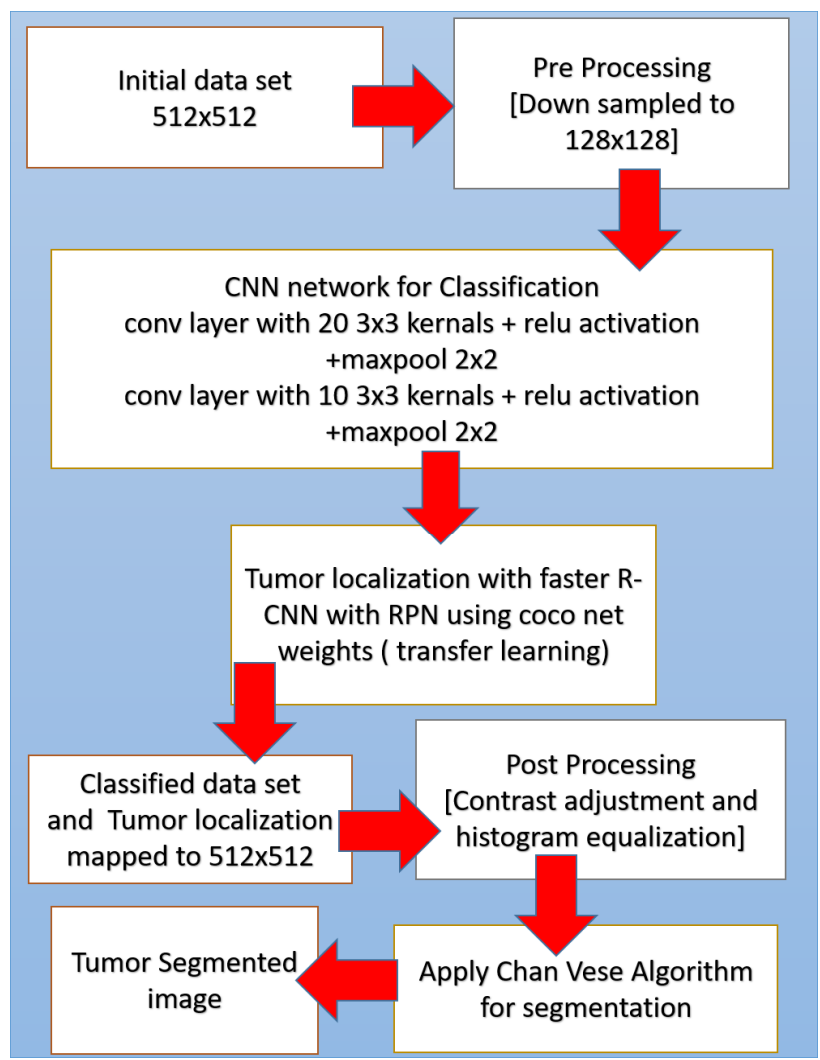

Figure 1: Proposed algorithm

Initially, a down sampled MR image is fed into a typical CNN to be classified as Meningioma or Glioma. Then at the second stage, the classified image goes through a faster RCNN networkfor tumor localization. The faster RCNN model adopted uses pre trained weights generated using COCO net data for its feature extracting CNN, which isfollowed by a Region Proposal network (RPN) and a classifier. The output of this second step is a boundary box around the tumor in 128*128 down sampled image. As the third step, this boundary box coordinates are mapped to 512*512 image with the original resolution, and Chan-Vese algorithm is applied only for the boundary box area. This approach assures that the Chan-Vesealgorithm converges to an accurate boundary. Hence, we were able to obtain a much precious boundary for tumor segmentation within a low computational time. The outputs obtained at the step 2 and 3 are presented in Figure 2.

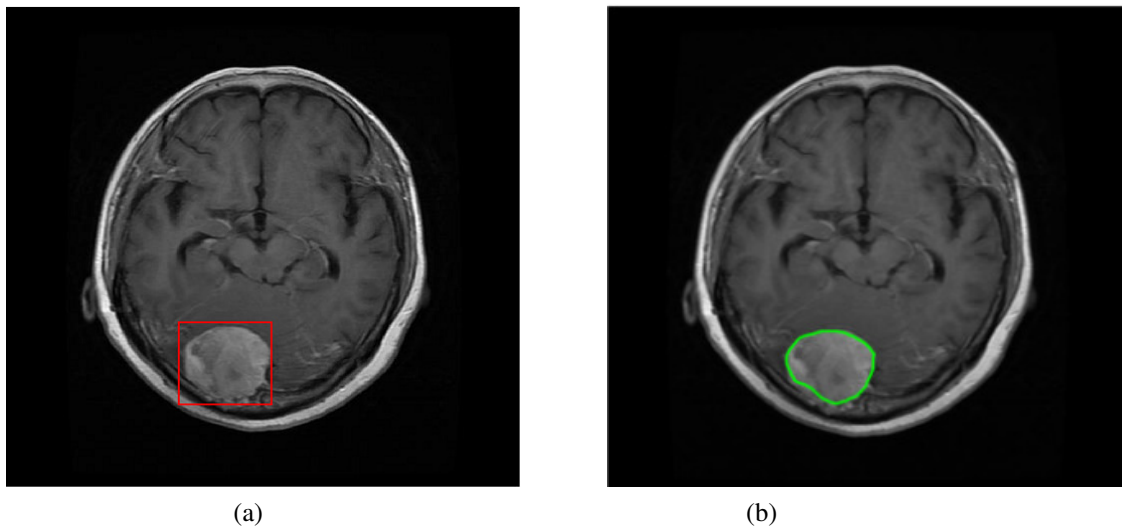

(a) (b)

Figure 2. (a)Brain tumor localization bounding box obtained at the step 2 (after Faster RCNN) (b) Tumor segmentation output obtained at the step 3 (after Chan-Vese).

## Background Works

### Basic Operation of CNN and R-CNN

CNN is a class of layered deep neural network architecture built using convolution, activation, pooling, and fully connected layers to analyse visual imagery. The convolutional layer uses set of learnable filters with different sizes to extract various feature maps to learn the correlation between neighbouring pixels, while drastically reducing the number of weighted parameters. The pooling layer introduce non-linear down sampling to the system architecture while the activation increase the nonlinear properties of the decision function of the overall network independent of the convolution layer. Followed by several combinations of convolution, pooling and activation, CNN has the fully connected layers, where high-level decision making takes place. At the final stage of the design dense layer or loss layer maps the trained outcome with the predefined output class. In a fully connected CNN architecture, these operations are executed forward and backward, through forward learning and back propagation as a designed architecture fine tune, i.e training cycle, to optimize the decision making capacity of the CNN architecture.

The R-CNN is an object detection and localisation mechanism evolved from CNN architecture. It is a region-based segmentation method which follows segmentation using recognition approach. It first extracts the free-form regions of interests from the input image and then conducts region-based classification on the extractedregion of interest(ROI). The faster R-CNN consists of two main subnetworks, R-CNN and RPN [27]. RPN itself narrows down the number of search regions in the image by generating anchors as in Figure 3 and works as a classifier that trains CNNs to classify, these selected ROIs, called hereafter as "regionproposals", into object categories. At first R-CNN takes an input image and segments it into many sub-images called regions with different dimensions. Next, each region is treated as an isolated image, and this isolated image is classified into predefined set of object labels. Finally a greedy algorithm is used to recursively combine sub-images with similar regions to generate the region proposals with the predicted object labels.

R-CNN use selective search algorithms to select these ROIs, that leads to a huge computational cost and slow response rate as it initially generates over 2000 regions for each input image. Hence, RPNbased bounding box detection algorithm was introduced into Faster R-CNN as the cost of generating region proposals is quite smaller in RPN compared to the selective search algorithm [27]. The significant difference between the two techniques is that the R-CNN uses the region proposals at the pixel level as input, whereas Faster R-CNN uses the region proposals at feature map level as its input. In general RPN generates 9 anchors using the input image as in Figure 4 and predicts the probability of an anchor being in the background or foreground. Based on two significant factors, positive or negative labels were assigned to these anchors. It is observed that anchors which has higher intersection-over-union (IOU), correspondence with the ground truth box. Hence, if an anchor and ground truth's IOU overlap is over 0.7, the anchor target gets a positive label, and if it is less than 0.3, the area is given a negative label [27]. The anchors where IOU lies between 0.3 and 0.7 (0.3<IOU<0.7) are not followed through for learning. The training phase of the RPN network is based on the loss function in (1), which is defined using the values assigned to the anchors.

$$L(\{p_i\},\{t_i\}) = \frac{1}{N_{cls}}\sum_i L_{cls}(p_i, p_i^*) + \lambda \frac{1}{N_{reg}}\sum_i p_i^* L_{reg}(t_i, t_i^*), \quad (1)$$

where, $i$ is the index of an anchor in a mini-batch, $p_i$ is the predicted probability of an anchor $i$ being an object, $t_i$ is the vector representing the 4 parameterized coordinates of the predicted bounding box, $L_{cls}$ is the classification loss (log loss over two classes), $L_{reg}$ is the regression loss, $N_{cls}$ is the mini-batch size and $N_{reg}$ is the number of the anchor locations.

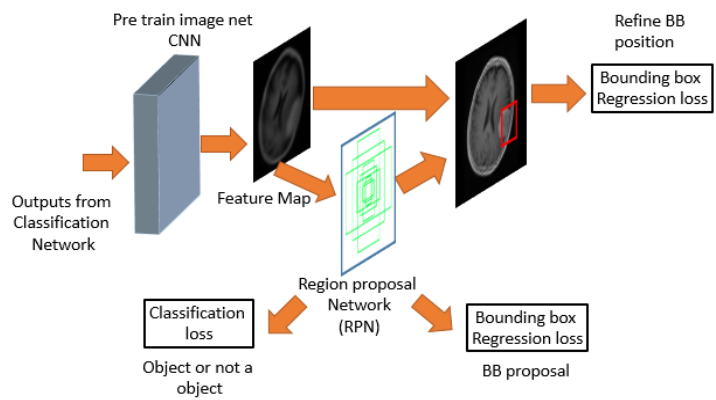

Figure 3: Region Proposal Network

It should be noted that when defining the loss function of the RPN for training purposes, a binary class label was assigned to each anchor. If the desired object isinside the anchor,algorithm assigns 1 for $p_i^*$ to indicatea positive anchor, whereas 0 is assigned to indicate a negative anchor.The $t_i^*$is that of the ground-truth box associated with a positive anchor. The regression loss, $L_{reg}$, of the loss function, adopted in faster R-CNN isdefined as,

$$L_{reg}(t_i, t_i^*) = R(t_i - t_i^*), \qquad (2)$$

where$R$ represents robust loss function. It should be noted that the regression loss is activated only for positive anchors ($p_i^*$= 1),and is disabled otherwise ($p_i^*$= 0), due to $p_i^* L_{reg}$ term in (1).

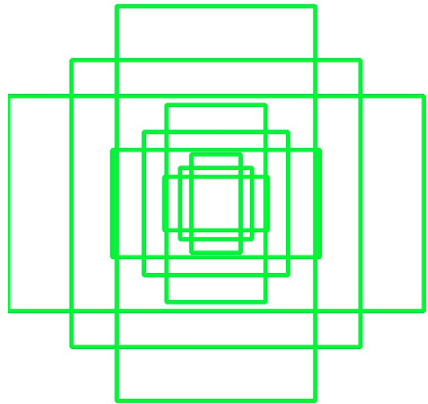

Figure 4: 9 Anchors used by RPN at (320,320)

A trained RPN generates different sizes of feature maps as its output. As it is not easy to work with different sizes of feature maps, ROI pooling splits the input feature map into equal size regions and applies max pooling to every region .It is worth noting that, output of the ROI pooling is always independent from the input size.

**Chan –Vese Segmentation**

In image processing, many edge detection techniques based on gradient of intensity, such as Sobel, Prewitt and Roberts, are used for object boundary detection and segmentation[23]. In MRI, it is challenging to get accurate boundary detection using these operators as MRI itself contains Rician noise, which causes irregularities in the edge estimation. As both Rician noise and the edges of the image contain high-frequency components, it is challenging to get accurate results with these edge detection operators which depend on the gradient of the intensities. In[28], although Gaussian smoothing filters were used to reduce the Rician noise, it resulted in blurred and distorted MRI, which may lead toerroneous diagnosis. Therefore it is worthwhile to exploit the edge detection algorithms which are not defined based on the gradient of the intensity for MRIs.

As an alternative to the edge based segmentation, active contour or Snake models [29] were developed for image segmentation which is governed by the gradient variation of the pixels. It starts with an initial estimation of theboundary curve plotted around the object of interest. With iterations,the estimated boundarymoves towards its interior and stops on the true boundary of the object based on an energy minimizing model [30]. Even though active contouring performs better than the operators that depend on the gradient of intensity, the final output is sensitive to initial condition of the algorithm. Hence, it is critical to set the correct boundary box at the beginning of the algorithm. Furthermore, the difficulties associated with topological changes like merging and splitting of the evaluation curve, contribute to active contour algorithm being less popular choice for segmentation problem.

Although there exist many algorithms with improvedsegmentation methodologies based on active contour algorithm, the level set method performs better with the noisy images[31][32]. A level set method is a powerful tool for contour evaluation which easily rein the topological changes like merging and splitting, which is difficult to tacklewith classical active contour models[30]. In literature, the level set function is denoted as $\emptyset(i,j,t)$, where $i,j$ are coordinates of the image and $t$ is the artificial time. Segmentation is defined as two regions where $\{\emptyset < 0\}$ belongs to region 1 and $\{\emptyset \geq 0\}$ belongs to region 2. The output of the level set function, edge contour, is defined in literatureusing $\{i,j\;;\emptyset(i,j,t) = 0\}$. This approach is used in Chan–Vese algorithm to form the level set function[30].

The Chan–Vese model, which is based on the Mumford–Shah functional for segmentation[14]is used to detect objects whose boundaries are not necessarily detectedby the gradient[33][30]. Mumford-Shah model is based on energy function and it finds a pair of $(u,C)$ for a given inputimage$u_0$. In here, $C$ denotes a smooth and closed segmentation curve of the object and $u$ represents nearly piecewise smooth approximation of the given image. Chan-Vesealgorithm presents an alternative solution to the Mumford-Shah model. It refers the fitting energy function and achieves the minimization through solving (3).

$$E^{CV}(c1,c2,C) = \mu.Length(C) + \lambda_1.\int_{inside(C)} |u_0(x,y) - c_1|^2\,dx\,dy + \lambda_2.\int_{outside(C)} |u_0(x,y) - c_2|^2\,dx \quad (3)$$

where $\mu$, $\lambda_1$ and $\lambda_2$ are positive constants. Typically $\lambda_1 = \lambda_2 = 1$, $c_1$ and $c_2$ are intensity average of given image $\mu_0$ inside $C$ and outside $C$ respectively. This minimization problem can be solved by replacing curve $C$ with level set function $\phi(x,y)$. If $\phi(x,y) > 0$ the point $(x,y)$ is inside the $C$, and if $\phi(x,y) < 0$ the point $(x,y)$ is outside $C$. In adition, $(x,y)$ is on the curve if $\phi(x,y) = 0$. For further details on Chan-Vese model reader is directed to[30][33][34][35]. The advantage of this method is that, even if the image is very noisy and initial conditions are not well defined, still the locations of the boundaries are accurately estimatedby the model.

## Materials and Methods

**Preparation of the dataset**

T1-weighed MRI brain tumor dataset presented at [13][36]is used in this research. It is a collection of MRI data from Nanfang Hospital, Guangzhou, China, and General Hospital,Tianjing Medical University, China from 2005 to 2010.It was first published online in 2015, and the last updated version was released in 2017. It has been extensively used in MRI tumor analysis research recently [16][18][37]. We have employed the updated version (in 2017) of the dataset for training, testing and validating of the proposed system. This dataset presents MRIs of coronal, sagittal and axial plan, of 233 patients with 3 types of brain tumors; namely meningioma (708 slices), glioma (1426 slices) and pituitary tumor (930 slices). The total images in the dataset are 3064 MRIs. Since this research only addressed the classification between meningioma and glioma, the pituitary tumor images were discarded at the pre-processing stage. Then the authors use only the axial MRI of meningioma (119 slices) and glioma (138 slices) tumorsfor the segmentation task. Hence, this research evaluates the performance of faster R-CNN under smaller annotated dataset environment.

**Pre-processing and the parameter setting**

At thepre-processing stage, the input images were resized into 128x128 as it was not feasible to train the neural network with the original size of 512*512.Yet, some finer features of the input present at original resolution, could be lost during the down sampling process, reducing the sensitivity of the output by a small fraction. Nevertheless, we have down sampled the input image to 128x128 to reduce the complexity of the network with the objective ofreducing the training of thenetwork.Initially, the dataset is randomly split into 3 sets as; training, testing and validation with the ratio of 0.70: 0.15: 0.15, and5fold cross-validation is applied to the training set with the scikit-learn library of python. Batch normalization was applied to input image to re-arrange the input intensities to the scale 0-1.

**Applying Chan-Vese Segmentation**

After extracting the bounding box through the proposed FasterR-CNN based model, a segmentation algorithm is applied to obtain a precise tumor boundary definition. As the first step of the segmentation process, the bounding box obtained at faster R-CNN from the down sampled 128*128 image was mapped to the original image size 512*512. Then general image pre-processing techniques such as contrast adjustment and histogram equalization were applied to adjust the contrast levels, brightness level, and the sharpness of the images, to reduce the noise levels while enhancing the details of the image. After applying the pre-processing techniques, Morphological Active Contours (Morphological Chan-Vese) technique was applied to identify the precise tumor region. In this research, Chan–Vese algorithm uses a square-level set function, instead of general circle-level set functions, because the input boundary boxes are defined as squares at the faster R-CNN.Furthermore the Parameters$\lambda_1$,and $\lambda_2$ in (3) were set to 1 following [30]. During the simulations100 iterations were followed to obtain the best convergence of the contour lines, and at each iteration, smoothing operators is applied 8 times. The values for these number of iteration and smoothing operator were selected by adopting trial and error method. After the segmentation, the output of the Chan-Vese algorithm is compared against segmentation output of Prewitt edge detection technique and ground truth demarcations provided by neurologists, to justify the performance of our proposed system.

**Statistical Performance Analysis**

For the performance evaluation of the overall cascaded MRI tumor segmentation system proposed, the ground truth demarcations provided by experts in the field (neurologists) were compared against the mask obtained from the prediction process of the designed system. Dice Score, Rand Index (RI), Variation of Information (VOI), Global Consistency Error (GCE), Boundary Displacement Error (BDE), Peak Signal to Noise Ratio (PSNR) and Mean absolute error (MAE) were calculated as the objective performance measure parameters [38].

Dice score (F1 score) is a statistical parameter which used to evaluate the similarity of two samples. Dice score lies between 0 and 1, with 1 signifying the highest similarity between predicted and truth. F1 score is calculated using,

$$\text{Dice score (F1 score)} = \frac{2\,TP}{2\,TP+FP+FN} \qquad (4)$$

where TP is true positive, FP is False positive, ,FN is False negative and TN is True negatives.

Rand index (RI) counts the fraction of pairs of pixels whose labelling is consistent between the computed segmentation and the ground truth image. RI lies between 0 and 1, and if the two images are identical, the RI should be equals to 1. The RI value is calculated using

$$RI = \frac{TP+TN}{TP+FP+FN+TN} \qquad (5)$$

VOI computes the measure of information content in each of the segmentations and how much of information one segmentation gives about the other, i.e it measures the information distance between the two segmentation. VOI is defined using the entropy and mutual information as,

$$VOI(S_g, S_t) = H(S_g) + H(S_t) - 2\,MI(S_g, S_t) \qquad (6)$$

where $S_g$ and $S_t$ are the fuzzy segmentations of the image, $H(S)$ is the marginal entropy, $H(S_g, S_t)$ is the joint entropy between two images, and $MI(S_g, S_t)$ is the Mutual information between two images[39].

The $GCE$ measures the extent to which, a particular segmentation can be viewed as a refinement of the other. Segmentations that are related are considered to be consistent, since they could represent the same image segmented at a different scale. The mathematical expression for GCE can be written as,

$$GCE(S_1, S_2) = \frac{1}{n}\min(\sum_i E(S_1, S_2, p_i), \sum_i E(S_2, S_1, p_i)) \qquad (14)$$

where $S_1$ and $S_2$ are two segmentations, and $p_i$ is a pixel position.

The boundary displacement error measures the average displacement error between the boundary pixels in the predicted segmentation and the corresponding closes boundary pixels in the ground truth segmentation as given bellow,

$$\mu_{LA}(u, v) = \{\frac{u-v}{L-1} \; ; \; 0 < u - v \quad (15)$$

Mean absolute error (MAE) is the average of the difference between predicted and the actual values in all test cases, i.e it is the average prediction error. Itis a quantity used to measure how close forecasts or predictions are to the eventual outcomes and the mathematical representation is given as,

$$MAE = \frac{1}{N}\sum_i \sum_j | E_{ij} - O_{ij} | \quad (16)$$

where N is the size of the image, E is the edge image, and O is the original image.

## Experimental Results

The experimental outcome of the implemented architecture presented in Figure 1, is analyzed at two stages; namely first after classification stage and second after segmentation stage. Both the training and validation accuracies of the classifier are presented using the confusion matrices in Figure 5.The ROC curves of the training and validationstages of the classification model are presented in Figure 6. Also, the performance of the segmentation algorithm is illustrated visually in Figure 7. Table2summarizes the statistical performance evaluations of the complete model for selected MRIs,whereas Table 3 presents the overall performance summery of the segmentation performance at the final stage of the proposed architecture.

**Results of classification model**

We present confusion matrix to illustrate the performance of the classification model against the ground truth. The Confusion matrices for the training dataset and the test samples are shown in the Figure 5.In each confusion matrix, green squares represents TP and TN values, light orange squaresrepresentsFP and FN values, and, blue squares were used to represents Positive Predictive Value(PPV), Negative Predictive Values (NPV), Specificity (Sp), and Sensitivity (Sn)respectively in clockwise, from the top right to bottom left. Overall correct classification rate (accuracy) was given in the purple square. The classification error for the trainingand testing set is equal to 7.69% and 6.42%.Overall summery of the classification process is tabulated in Table 1.

As the dataset is more biased towards glioma, we have used Cohen's kappa statistic[40]as itis a very good measure of the performance of a classifier against the imbalanced class problem. It estimates the designed classifier performance against a random guessing classifier based on the frequency of the class occurrence. In general,Cohen's kappa value above 0.81 is an indication of a perfect classifier while value less than 0 indicates a non performing classification outcome.The Cohen's kappa statistics value for the proposed model is of 0.843 in training and 0.872 in testing which indicatesagoodagreementin the classification process.

Area Under the Curve -Receiver Operating Characteristic (AUC-ROC) isanother powerful metrics used to evaluate the performance of machine learning algorithms with imbalance dataset. ROCcurveillustratesTPrate versus FPrate at various thresholdvalues and is commonly used in medical statistics.The ROC curve for training model and testing are shown in Figure 6 and AUC values are at 0.93 and 0.94 respectively.

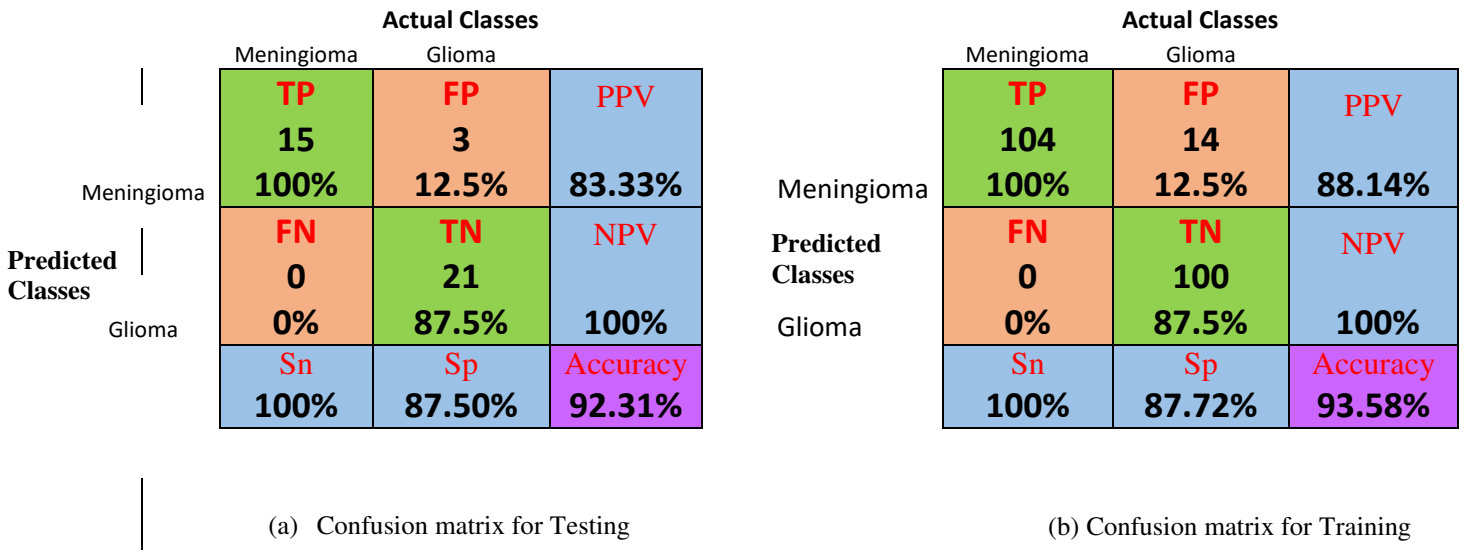

(a) Confusion matrix for Testing　　　　　　　　　　　　(b) Confusion matrix for Training

Figure 5: Confusion matrices

Table 1:Performance of the classification network

| Brain tumor class | Glioma | Meningioma |
|---|---|---|
| No if images taken for transfer learning | 114 | 104 |
| Training accuracy | 88.5% | 100% |
| Testing accuracy | 87.5% | 100% |
| No if images taken for testing | 24 | 15 |
| No of correctly predicted images | 21 | 15 |

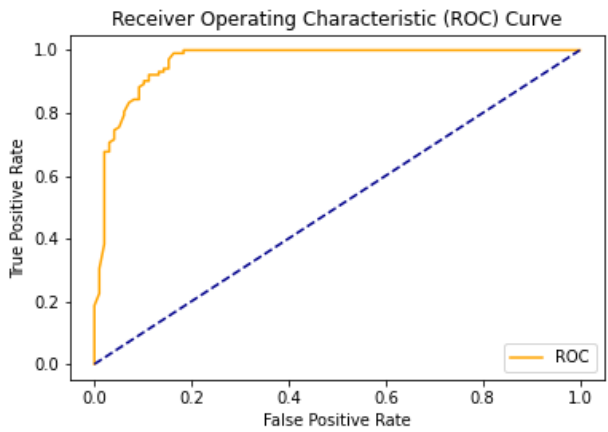 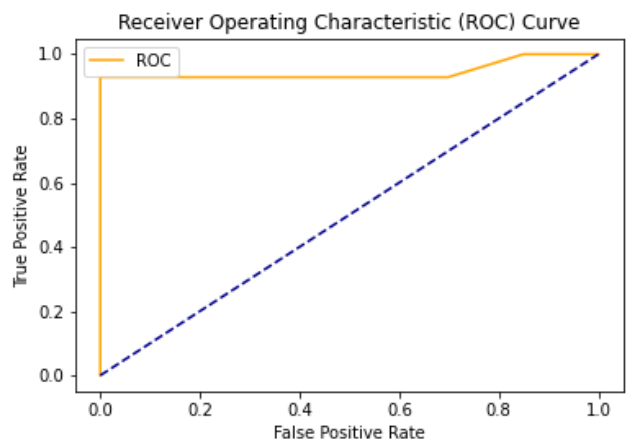

(a)Training　　　　　　　　　　　　　　　　　　　　(b) Testing

Figure 6: Receiver Operation Characteristic (ROC) Curve

**Results of segmentation model**

In our proposed model, the Faster R-CNN extracts the bounding box of the tumor and it is followed by Chan-Vese algorithm to obtain a precise tumorboundary for an accurate segmentation of the tumors. The Faster R-CNN model was able to generatethe boundary boxes with 93.6%confidence intervaland 99.81% accuracy. Figure 7 illustrates random sample outputs of the faster R-CNN.As in Figure 7, (a), thetumoris localized correctly with 99% confidence interval. There are two possible outcomes predicted in Figure 7(b). Once closely examined, it is evident that the false positive area, which comprises of brain matter, is predicted as tumour only with 50% confidence interval. The correcttumorarea in Figure 7. (b) is identified with 99% confidence interval.

However, in certain slices of axial MRIs where the image consists of complex anatomical structures such as skull and the eye socket, a false detection could happen with high confidence interval (Example: Tumor 5 of table 2). In such scenarios the system fails to completely recover from the erroneous detection. Yet, close examination of the results presented in table 2, yields that most accurate detections have a confidence level greater than 90%, compared to the false detections which is of less than 80%. Another false detection is presented in Figure 7. (c), where the false detection carries 80% of confidence level,while all the other positive detections carry confidence level of 98%

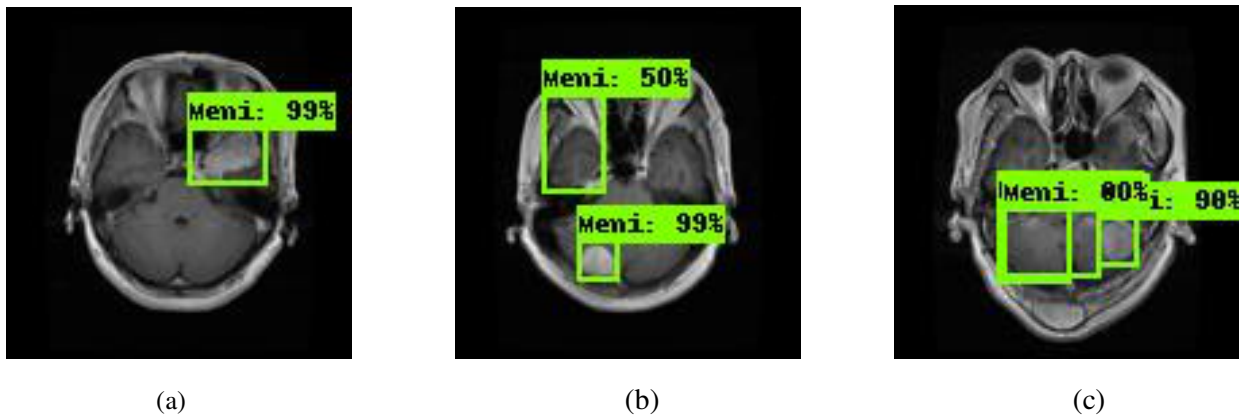

(a)        (b)        (c)

Figure 7: Random Sampled of Tumor localization of the faster RCNN network

Table 2: Performance comparison between Chan - Vese (CV) vs Prewitt for selected tumorMRIs

| Sample MRIs | 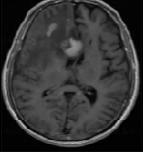 Tumor 1 | | 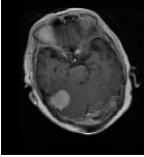 Tumor 2 | | 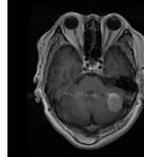 Tumor 3 | | 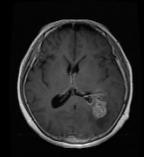 Tumor 4 | | 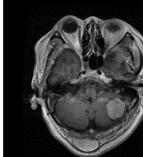 Tumor 5 | |
|---|---|---|---|---|---|---|---|---|---|---|
| Confidence level of detection | 0.99 | | 0.98 | | 0.98 | | 0.99 | | Correct detection with 80%, Errornous diagnosis 98%. | |
| Boundary outlining method | CV | prewitt | CV | prewitt | CV | prewitt | CV | prewitt | CV | prewitt |
| Dice score | 0.9256 | 0.9025 | 0.9125 | 0.8545 | 0.9237 | 0.8723 | 0.9164 | 0.8845 | 0.6256 | 0.5941 |
| Accuracy | 0.9427 | 0.9254 | 0.9485 | 0.9123 | 0.9486 | 0.9406 | 0.9503 | 0.9372 | 0.7831 | 0.7232 |
| RI | 0.9872 | 0.9542 | 0.9953 | 0.9648 | 0.9967 | 0.9865 | 0.9955 | 0.9947 | 0.8263 | 0.7639 |
| VOI | 0.0112 | 0.1626 | 0.0252 | 0.2683 | 0.0281 | 0.0965 | 0.0376 | 0.0752 | 0.873 | 1.94 |
| GCE | 0.0034 | 0.1795 | 0.005 | 0.068 | 0.0031 | 0.046 | 0.0044 | 0.0089 | 0.7361 | 0.927 |
| BDE | 1.83 | 2.57 | 2.0298 | 3.325 | 1.3171 | 2.56 | 1.7234 | 2.78 | 5.8263 | 7.7836 |
| PSNR | 30.5287 | 13.18 | 27.2865 | 15.26 | 27.8105 | 9.56 | 26.499 | 16.25 | 13.578 | 9.354 |
| MAE | 58.54 | 143.23 | 51.28 | 121.56 | 65.23 | 128.26 | 45.6 | 99.57 | 86.6521 | 145.764 |

To determine the overall performance of the proposed Chan-Vese algorithm based system, we compare the output of Chan-Vese algorithm against the ground truth, gold standard, as well as against the output of a simple gradient based edge detection technique. Prewitt is adopted in the experiment as the gradient based edge detection method. The statisticalquality measures obtained for the segmented output of both the Chan-Vesealgorithm, and the Prewitt edge detection, are tabulated in Table 2. Columns with Tumors 1-4 presents positive detections, while column with Tumor 5 presents a false detection. The value of GCE, VOI and BDE must be low, whereas the RI should be high for a good segmented image. It is observed from the results that the Chan-Vese algorithm exhibits a superior performance than the gradient-based edge detection algorithm, Prewitt. This is also evident from the visual inspections shown in Figure 8.Although the RI for both Chan-Vese and Prewitt algorithms have a significantly higher score for all the test images, the RI values of Chan-Vese algorithm have a relatively higher value than the Prewitt. Also, the MAE value, which should be low for better prediction, is lowest for Chan-Vese according to Table 2. Hence we can conclude that the tumor boundary detected by Chan-Vese is superior than that of Perwitt

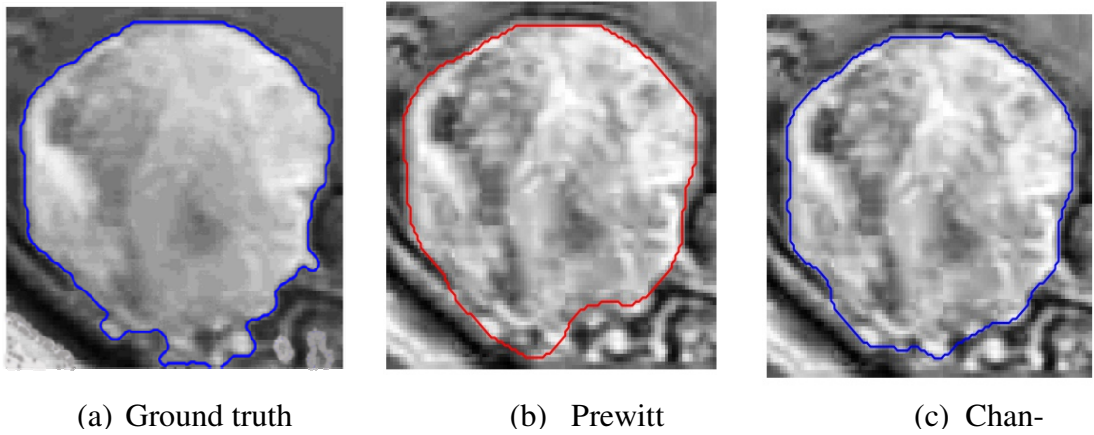

(a) Ground truth       (b) Prewitt       (c) Chan-

Figure 87: Segmented images vs Ground truth

Table 3: Overall performance of the system

| Boundary estimation method | Chan-Vese | Prewitt |
| --- | --- | --- |
| Dice score | 0.92 | 0.90 |
| Accuracy | 0.9457 | 0.9125 |
| RI | 0.9936 | 0.9642 |
| VOI | 0.0301 | 0.1542 |
| GCE | 0.004 | 0.083 |
| BDE | 2.099 | 3.581 |
| PSNR | 77.076 | 13.43 |
| MAE | 52.946 | 126.47 |

However, in some critical casesas illustrated in Figure 7 (c), the system may not perform as desired. Furthermore, the performance measurements for such a scenario is presented in the Last column of Table 2, Tumor 5. It is evident that some of the performance measures for false detection, specifically GCE, VOI and BDE, shows a deviation from that of positive detection.
The main reason behind false detection is the miss-prediction and classification of brain matters as tumors. These inaccuracies in the model can be easily rectified by further tuning the proposed model using a larger dataset.

The objective of the system presented in this manuscript is to obtain a finer segmentation of brain tumors. To achieve this objective, we firstly employed Faster R-CNN to obtain the initial boundary region and secondly used Chan-Veseactive contour to refine initial boundary assessment to obtain an exact boundary extraction.  Table 3 summarise the overall performance of the proposed architecture. According to the objective measurements, the overall system achieved a Dice Score of 0.92, accuracy of 0.946, RI of 0.99 and PSNR of 77.1, against the gold standard,pointing to excellent precision of the proposed segmentation method.

# Discussion

This study presents an automated MRI tumor classification and segmentation algorithm based on deep learning techniques and active contours. Results obtained from the experiments demonstrate remarkable performance at braintumor segmentation with dice score of 0.92, accuracy of 0.9457, RIof 0.9936, VOI of 0.0301, GCE of 0.004, BDE of 2.099, PSNR of 77.076 and MAE of 52.946.

The model presented in this manuscriptprovides 0.915 dice score for glioma segmentation with thefigshare data set [13][11], with the faster R-CNN and Chan-Vesealgorithms. In comparison, authors in [41]has developed a CNN based glioma segmentation algorithm and achieved a 0.87 Dice score on BRATS 2013 and 2015 datasets. Experimental results were presented in [20] for accurate glioma segmentation algorithm which obtained 0.897 Dice score. An automatic semantic segmentation model was developed on the BRATS 2013 dataset by the authors in [42] and the Dice score was around 0.80.

In our study, 0.926 dice core was obtained only for the meningioma segmentation which is comparably high amongsimilar research works. Authors at [43] obtained dice coefficient around 0.81 for 56 meningioma cases by using deep learning on routine multi-parametric model. One recent study [44] achieved dice coefficients ranging between 0.82–0.91 by employing an algorithm based on 3D deep convolutional neural network.

Furthermore authors in [45] proposed a CNN based algorithm on same figsharedataset [13] and achieved a dice score of 0.71 for both meningioma and glioma together with axial MR images. Authors in [37] were able to increase the segmentation accuracy using Cascaded Deep Neural Networks and obtained around 0.8 in dice score in both Meningioma and Glioma segmentation. In our study, we achieved an average dice score of 0.92 for both meningioma and glioma segmentation using the same dataset.

Hence, comparison with the comparable state-of-the-art methods shows that the proposed methodology exhibits a remarkable improvement in brain tumor segmentation. Nevertheless, all these researchers proved that deep neural networks are capable of performing significantly accurate brain tumor segmentation in MR images. We show that the segmentation output can be further improved by using Active contour algorithms along with deep learning architectures.

# Conclusions

In the research presented, we have proposed a R-CNN and Chan-Vese algorithms based model for meningioma and glioma brain tumor classification and segmentation. The proposed model is validated using figshare dataset with 5-fold cross-validation and objective quality metrics dice score, RI, VOI, GCE, BDE, PSNR and MAE are calculated to analyse the performance of the proposed architecture. We have used R-CNN to obtain the initial tumor boundary box which is followed by active contouring to obtain exact tumoroutline. We adopt level set functions based, Chan-Vese algorithm which is independent of Rician noise, for both meningioma and glioma brain tumor segmentation and we compare the performance of the proposed segmentation method against that of the typical-gradient based edge detection algorithm Prewitt. We were able to achieve much more accurate segmentation result through the Chan-Vese algorithm with an average dice score of 0.92 for both tumor types.In addition experts in the field have cross checked our segmentation output and have validated with a high confidence interval. In the research presented, we have shown with evidence that active contour algorithms along with localized outputs of deep learning architectures such as R-CNN, is capable of improving the segmentation accuracy and the precision in MRI tumor segmentation applications. Hence we can conclude that the model present can be used as a reliable aid for brain tumor classification and segmentation in low human resource, expertise, environments.

## Data Availability

All MRI data are available in the brain tumor imaging archive (https://figshare.com/articles/brain_tumor_dataset/1512427)

## Conflicts of Interest

The authors declare that there is no conflict of interest.

## Acknowledgments

The authors wish to acknowledge the insightful and extremely helpful reviews and comments provide by Dr. S.C Weerasinghe, Neurologist, atTeaching hospital Anuradhapura during the design and validation stages of the research.

onfigs. [Accessed: 05-Oct-2020].

# Appendix

**Proposed algorithm**

The first stage of the proposed system is the classification, where input MRI images are analysed to detect the presence and absence of the tumours. To achieve this objective, a simple CNN architecture with total of 5 layers was designed. The CNN network designed consist of two 2D convolutional (2DConv) layer with 20-$3x3$ kernals and 10-$3x3$ kernels respectively.Both of the 2DConv layers adopts non-linearity through, Relu activation functionand each layer is followed by maxpooling layers with $2x2$ window size. The 2D feature map generated through convolution and maxpooling is converted into 1D feature vectorsby flattening and these 1D feature vectors are fed into the fully-connected layer for the classification task.The output layer classifies the inputs as a meningioma or glioma using softmaxfunction. Adam optimizer was used as the optimization function of the convolution layers and learning rate was set to 0.02 in the training process. The proposed classification architecture was trained and validated, before testing for the classification accuracy. The validation process was carried out by using 5 fold cross validation with 500 epochs at each validation fold.At the end of first stage the system is capable of classifying the input MRI according to the tumor type.

Next, the classified images were fed into the second stage of the proposed architecture, which is the Region Proposal Network (RPN)based faster R-CNN for tumor localization. Due to the limitation of annotated images in the dataset, transfer learning method with pre trained faster R-CNN weights[46]was utilized in our architecture. The faster R-CNN was able to localize the tumor with a bounding box and produce a confidence interval for each decision.

As the last part of the proposed architecture, the Chan-Vese active contour algorithm was used to segment the tumor in the Axial MRI. In general, active contour algorithms are utilized to obtain finer and precise object boundaries, given a user defined initial guessed boundary. Thus, in MRI brain tumor segmentation using Chan-Vese algorithm, it is essential to provide a reasonably accurate initial region of interest searcharea for successful segmentation.We have used the bounding box obtained after faster R-CNN as the initial boundary for Chan-Vese algorithm to automate the entire brain tumor segmentation process.